\documentclass[10pt,english,english,nofootinbib,notitlepage,superscriptaddress,aps,prd,twocolumn]{revtex4-1}

\usepackage{enumerate}
\usepackage{amsmath}
\usepackage{amsfonts}
\usepackage{amssymb}
\usepackage[utf8]{inputenc}
\usepackage[T1]{fontenc}
\usepackage{mathtools}
\usepackage{wasysym}
\usepackage{accents}
\usepackage{float}
\usepackage[svgnames]{xcolor}
\usepackage[colorlinks=true, pdfstartview=FitV, linkcolor=blue, citecolor=red, urlcolor=magenta]{hyperref}
\usepackage{url}
\usepackage{graphicx}
\usepackage{subfig}
\usepackage{bm}
\usepackage{tikz}
\usetikzlibrary{arrows,decorations.markings}
\usepackage{csquotes}

\usepackage{changepage}

\begin{document}
\newcommand{\Areia}{
\affiliation{Department of Chemistry and Physics, Federal University of Para\'iba, Rodovia BR 079 - Km 12, 58397-000 Areia-PB,  Brazil.}
}
\newcommand{\Lavras}{
\affiliation{Physics Department, Federal University of Lavras, Caixa Postal 3037, 37200-000 Lavras-MG, Brazil.}
}

\title{Thermal Quantum Correlations in Two Gravitational Cat States}

\author{Moises Rojas}\email{moises.leyva@ufla.br}
\Lavras

\author{Iarley P. Lobo}\email{lobofisica@gmail.com}
\Areia
\Lavras


\begin{abstract}
We consider the effect of a thermal bath on quantum correlations induced by the gravitational interaction in the weak field limit between two massive cat states, called gravitational cat (gravcat) states. The main goal of this paper is to provide a good understanding of the effects of temperature and several parameters in the entanglement (measured by the concurrence) and quantum coherence (measured by the $l_1$-norm that is defined from the minimal distance between the quantum state and the set of incoherent states) which are derived from the thermal quantum density operator. Our results show that the thermal concurrence and  $l_1$-norm can be significantly optimized by increasing the masses or decreasing the distance between them. We investigate and discuss the behavior of these quantities under temperature variations in different regimes, including some that are expected to be experimentally feasible in the future. In particular, we observe that thermal fluctuations raise non-entangled quantum correlations when entanglement suddenly drops.
\end{abstract}

\pacs{}
\maketitle


The quantization of the gravitational interaction has occupied a prominent space in the research of gravity and quantum field theory in the past several decades. This quest has been mainly focused on efforts to coherently quantize the gravitational field or the spacetime geometry \cite{Kiefer:2007ria,Rovelli:1997yv,Mukhi:2011zz,Ambjorn:2010rx}. Since the beginning of the 21st century, the research in quantum gravity has entered a new era in which phenomenological approaches to this problem have arisen due to technological advances that have allowed experimental bounds on predictions with Planck scale sensitivity \cite{AmelinoCamelia:2008qg,Addazi:2021xuf}. Besides constraints from observations of astrophysical events \cite{Acciari:2020kpi,Addazi:2018uhd}, underground and laboratory-based measurements have been considered in recent investigations (see, for instance, \cite{Arzano:2019toz,Lobo:2020qoa,Carney:2018ofe,Addazi:2017bbg}.)
\par
Recently, a more fundamental question regarding this problem has been matter of debate: how can one verify if the gravitational interaction indeed needs to be quantized \cite{Marletto:1}? Besides that, is there a model-independent experimental protocol that can assure whether gravity is an interaction of quantum mechanical nature \cite{Marletto:2017kzi,Bose:2017nin}? In these approaches, based on the theory of quantum information, it is argued that an interaction that is responsible for creating entanglement between two systems must be of quantum nature. Therefore, sufficient evidence of quantum gravity would be the detection of entanglement between massive states induced by the gravitational interaction \cite{Rovelli:2021thx,Bose:2022uxe}.
\par
Despite the ongoing debate on the matter \cite{Christodoulou:2018cmk,Christodoulou:2018xiy,Hall:2017nzl,Reginatto:2018ksa,Anastopoulos:2020cdp}, the existence of entanglement (besides other correlations) between massive states constitutes an exciting topic to be analysed by itself. In this paper, we do not aim to discuss the issue of whether the detection of gravity-induced entanglement represents a test of the quantum nature of the gravitational interaction; instead, we pose a very simple question that we believe has not been sufficiently addressed in the literature of this specific topic: how are the entanglement and other quantum correlations in this scenario affected by a thermal bath? The different subject of some environmental effects in this setup has been analyzed previously for the case of the decoherence time necessary to observe entanglement \cite{Rijavec:2020qxd,DiBartolomeo:2021jpc}, and its robustness against stochastic fluctuations was verified \cite{Nguyen:2019huk}, besides other recent investigations on this matter \cite{Miki:2020hvg,Qvarfort:2018uag}. However, here we aim to analyze how the influence of temperature (through a thermal bath) might destroy some correlations presented in this context.
\par
The quantum \textit{entanglement} plays an important role in quantum computation and quantum information \cite{Bennett:1992tv,Bennett:1995tk,Amico:2007ag}, which is a tool that has been recently explored for simulating properties of a quantum spacetime \cite{Li:2017gvt,Czelusta:2020ryq}. In a similar manner, \textit{quantum coherence} \cite{l1norm1,Streltsov:2016iow} is also a central concept in quantum mechanics and has been extensively explored in the context of quantum information processing \cite{Streltsov:2016iow}, quantum metrology \cite{frowis,giovannetti}, thermodynamics \cite{santos,lostaglio}, {effects of a minimum length scale through the use of a Generalized Uncertainty Principle \cite{Petruzziello:2020wkd} or the assumption of modified spacetime symmetries in the Schr\"{o}dinger equation \cite{Arzano:2022nlo}}, among others. 
\par
Moreover, the quantum coherence is a fundamental measure of the quantum nature of a given phenomenon; for instance, it governs effects such as interference of states and provides important tools for quantum optics \cite{Glauber:1963tx,Sudarshan:1963ts}. Besides that, it has been recently shown a close relation between quantifiers of coherence and entanglement \cite{Streltsov:2015xia}. In recent years, employing quantum coherence as a resource, several interesting results have been obtained in different quantum systems, among them, in quantum thermodynamics \cite{Kammerlander,conditionalentropy}, quantum dots \cite{q-dots}, Heisenberg spin chains \cite{spin-chains}, spin waves \cite{steady}, quantum batteries \cite{batteries}, quantum computation \cite{shor,envir}, among others.
\par 
Several quantities have been proposed as candidate measures of quantum coherence. In particular, Baumgratz et al. \cite{l1norm1} introduced the $l_{1}$-norm of coherence and the relative entropy of coherence as measures of this property. In the present work, we confine ourselves to utilize the $l_{1}$-norm of coherence.
In this paper, we aim to contribute to the research on the use of these quantum correlations in gravity, by investigating the way temperature (induced by a thermal bath) impacts quantifiers of entanglement and quantum coherence. We stress that we will consider a closed system, but in thermal contact with a thermal bath. Further analyses regarding the assumption of an open system and indeed environmental effects \cite{environment1,environment2} go beyond the scope of this paper.
\par
The present paper is structured as follows: In Section \ref{sec:model}, we present the model describing the gravitational interaction that generates two qubits (which we call gravcat states), and we introduce the thermal density operator due to the thermal bath. Subsequently, in Section \ref{sec:thermalqc}, we present thermal concurrence as a quantifier of entanglement and the $l_1$-norm as a quantifier of quantum coherence (both depend on the temperature of the thermal bath). In Section \ref{sec:results}, we present the evolution of these quantum correlations as a function of the temperature and the different parameters of the model, in which the temperature range for which the system remains entangled is shown, and the appearance of thermal fluctuations is discussed. Finally, in Section \ref{sec:conc}, our conclusions and further discussions are presented.


\section{The Model}\label{sec:model}
The model consists of a set of two particles of mass $m$, each one in a one-dimensional double-well potential, with local minima at $x=\pm L/2$, which we will call gravitational cats (gravcats). The potential is even and is such that we can associate two eigenstates for each particle describing its localization in each of the minima $|\pm\rangle$, where $\hat{x}|\pm\rangle=\pm \frac{L}{2}|\pm\rangle$. From the Landau--Lifschitz approximation \cite{Anastopoulos:2020cdp}, they can be written in terms of the ground $|0\rangle$ and first excited states $|1\rangle$ as
\begin{equation}
    |\pm\rangle=\frac{1}{\sqrt{2}}(|0\rangle\pm |1\rangle).
\end{equation}
\par
The use of just two states is justified by the fact that, in this model (described in details in \cite{Anastopoulos:2020cdp}), the energy difference between the ground and first excited states (calculated using the WKB method in section 3 of \cite{Anastopoulos:2020cdp}) is much smaller than the energy of the second excited state. This allows such two-dimensional description.
\par
The experimental setup that could be modeled as above can be found in some references in this subject, for instance \cite{Marletto:2017kzi,Bose:2017nin,Carlesso:2019cuh}, such as massive molecules, nanomechanical oscillators, diamond microspheres, optomechanical systems, etc.
\par
\begin{figure}[H]
\centering
\includegraphics[scale=0.4]{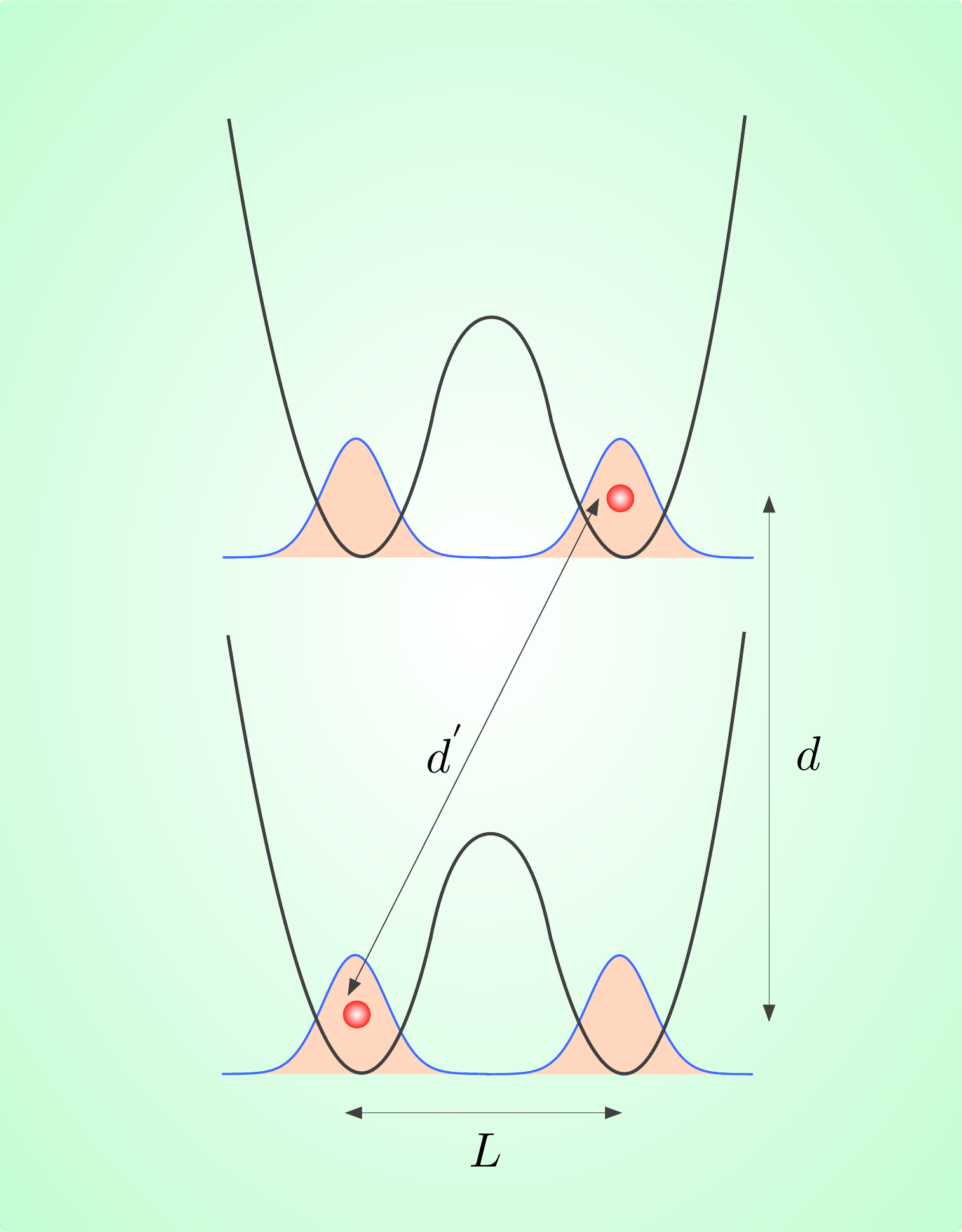}\caption{\label{fig:GCat}(Color online) Schematic representation of our model. Two particles are located in an even double well potential, whose minima are separated by a distance $L$. The quantity $d$ ($d'$) describes the distance between the particles when each of them are at the same (different) relative minima.}
\end{figure}

The Hamiltonian $\mathcal{H}$ for this model has been analyzed in \cite{Anastopoulos:2020cdp} and can be written as
\begin{eqnarray}
\mathcal{H}= & \frac{w}{2}\left(\sigma_{z}\otimes\mathbb{I}+\mathbb{I}\otimes\sigma_{z}\right)-\Delta(\sigma_{x}\otimes\sigma_{x})\, ,\label{eq:hamiltonian}
\end{eqnarray}
where $\sigma_{x,z}$ is usual Pauli matrices; $w$ is energy difference between the ground and first excited states and furnishes the energy scale of this setup.\footnote{We assume $\hbar=1$.} We also have a quantity $\Delta$ that measures the intensity of the gravitational interaction between the states with masses $m$, and is given by $\Delta=\frac{\alpha}{2}\left(\frac{1}{d}-\frac{1}{d^{'}}\right)$, where $\alpha=Gm^{2}$, $G$ is a universal gravitational
constant; besides that, $d$ and $d'$ are the relative distances between the two masses, as depicted in Figure \ref{fig:GCat}. 

After the diagonalization
of the Hamiltonian, we obtain the following eigenvalues: 
\begin{eqnarray}
\varepsilon_{1}= & -\Delta,\nonumber \\
\varepsilon_{2}= & \Omega,\nonumber \\
\varepsilon_{3}= & -\Omega,\nonumber \\
\varepsilon_{4}= & \Delta\, ,\label{eq:eigenvals}
\end{eqnarray}
where $\Omega^{2}=\Delta^{2}+w^{2}$. Their corresponding eigenstates
are described in terms of the standard basis, respectively, by
\begin{eqnarray}
|\varphi_{1}\rangle= & \frac{1}{\sqrt{2}}\left(|10\rangle+|01\rangle\right),\\
|\varphi_{2}\rangle= & \sin(\theta_{-})|11\rangle+\cos(\theta_{-})|00\rangle,\label{eq:phi2}\\
|\varphi_{3}\rangle= & \sin(\theta_{+})|11\rangle+\cos(\theta_{+})|00\rangle,\label{eq:phi3}\\
|\varphi_{4}\rangle= & \frac{1}{\sqrt{2}}\left(-|10\rangle+|01\rangle\right),
\end{eqnarray}
where $\theta_{\pm}=\arctan \left(\frac{\Delta}{w\pm\Omega}\right)$.


\subsection*{Thermal Density Operator}

In order to study the thermal effects on the quantum correlations
of our model, we consider the system in thermal bath, the system state
in the thermal equilibrium is described by $\rho(T)=\frac{\exp\left(-\beta\mathcal{H}\right)}{Z}$,
where $\beta=\frac{1}{k_{B}T}$, with $k_{B}$ being the Boltzmann's
constant, $T$ is absolute temperature, and the partition function of the system is defined by $Z=Tr\left[\exp\left(-\beta\mathcal{H}\right)\right]$. The thermal state of this system can be described by the quantum density operator

\begin{equation}
\rho=\left[\begin{array}{cccc}
\rho_{1,1} & 0 & 0 & \rho_{1,4}\\
0 & \rho_{2,2} & \rho_{2,3} & 0\\
0 & \rho_{2,3} & \rho_{2,2} & 0\\
\rho_{1,4} & 0 & 0 & \rho_{4,4}
\end{array}\right],\label{eq:rho-mat}
\end{equation}
whose elements are expressed by

\begin{eqnarray}
\rho_{1,1}= & \frac{1}{Z}\left(\mathrm{e}^{-\beta\varepsilon_{2}}\sin^{2}{(\theta_{-})}+\mathrm{e}^{-\beta\varepsilon_{3}}\sin^{2}{(\theta_{+})}\right),\nonumber \\
\rho_{1,4}= & \frac{1}{Z}\left(\frac{\mathrm{e}^{-\beta\varepsilon_{2}}\sin{(2\theta_{-})}+\mathrm{e}^{-\beta\varepsilon_{3}}\sin{(2\theta_{+})}}{2}\right),\nonumber \\
\rho_{2,2}= & \frac{1}{Z}\left(\frac{\mathrm{e}^{-\beta\varepsilon_{1}}+\mathrm{e}^{-\beta\varepsilon_{4}}}{2}\right),\nonumber \\
\rho_{2,3}= & \frac{1}{Z}\left(\frac{\mathrm{e}^{-\beta\varepsilon_{1}}-\mathrm{e}^{-\beta\varepsilon_{4}}}{2}\right),\nonumber \\
\rho_{4,4}= & \frac{1}{Z}\left(\mathrm{e}^{-\beta\varepsilon_{2}}\cos^{2}{(\theta_{-})}+\mathrm{e}^{-\beta\varepsilon_{3}}\cos^{2}{(\theta_{+})}\right),
\end{eqnarray}
where $\varepsilon_k$ are the eigenenergies of the Hamiltonian ${\cal H}$, given by the energies of the eigenstates $\varphi_k$, and $Z=Tr\left[\exp\left(-\beta\mathcal{H}\right)\right]=\sum_i \exp(-\beta \varepsilon_i)$ assumes this form for the diagonalized Hamiltonian. The thermal density operator assumes such $X$-shape form due to the symmetry of the Hamiltonian, in addition to the
thermal density operator must be Hermitian, also $Tr(\rho)=1$, $\rho^{2}\neq\rho$ and $Tr(\rho^{2})\leq1$,
due to the fact that we are dealing with a mixed density operator. Since $\rho(T)$ describes a thermal state, correlations of this system are called \textit{thermal quantum correlations}.


\section{Thermal Quantum Correlations}\label{sec:thermalqc}

In this section, we discuss the thermal entanglement and quantum coherence
for two gravcats in contact with a thermal bath.

\subsection{Thermal Entanglement}

To describe the thermal entanglement of the state $\rho$, we use the concurrence defined by Wootters \cite{wootters,hill},
which is given by
\begin{equation}
\mathcal{C}=\mathrm{max}\{\sqrt{\lambda_{1}}-\sqrt{\lambda_{2}}-\sqrt{\lambda_{3}}-\sqrt{\lambda_{4}},0\},\label{eq:Cdf}
\end{equation}
assuming $\lambda_{i}$ are the eigenvalues in decreasing order of
the matrix
\begin{equation}
R=\rho\left(\sigma_{y}\otimes\sigma_{y}\right)\rho^{*}\left(\sigma_{y}\otimes\sigma_{y}\right),\label{eq:R}
\end{equation}
where $\rho^{*}$ denotes the complex conjugation of $\rho$ and $\sigma_{y}$ is a Pauli matrix.

In our model, the concurrence of the state can be obtained from Equation (\ref{eq:Cdf}), which results in 
\begin{equation}
\mathcal{C}=2\mathrm{max}\left\{ \mid\rho_{2,3}\mid-\sqrt{\rho_{1,1}\rho_{4,4}},\mid\rho_{1,4}\mid-\mid\rho_{2,2}\mid,0\right\}.\label{eq:conc}
\end{equation}


\subsection{Quantum Coherence}

It is well known that quantum coherence is an important physical resource in different quantum information processing tasks.  An intuitive and computable measure of quantum coherence is the $l_{1}$-norm of coherence \cite{l1norm1}. This measure quantifies coherence by using the minimal distance between the quantum state and the set of incoherent states.
The $l_{1}$-norm coherence of the quantum state $\rho=\sum_{i,j}\rho_{i,j}|i\rangle \langle j|$ is the sum of the absolute values of all off-diagonal elements entries
\begin{equation}
\mathcal{C}_{l_{1}}(\rho)=\sum_{i\neq j}|\rho_{i,j}|.\label{eq:norm-l1}
\end{equation}
Calculating the $l_{1}$-norm for the two gravitational cat states is straightforward and results in
\begin{equation}
\mathcal{C}_{l_{1}}(\rho)= 2|\rho_{2,3}|+2|\rho_{1,4}|.\label{eq:norm}
\end{equation}

In the next section, we review and discuss the main results.

\section{Results and Discussion}\label{sec:results}
In what follows, we focus on the thermal quantum correlations of two gravitational cat states with gravitational interaction between the states. In what follows, unless the opposite is stated, we shall consider $k_B=1$.

\begin{figure}[H]
\centering
\includegraphics[scale=0.4]{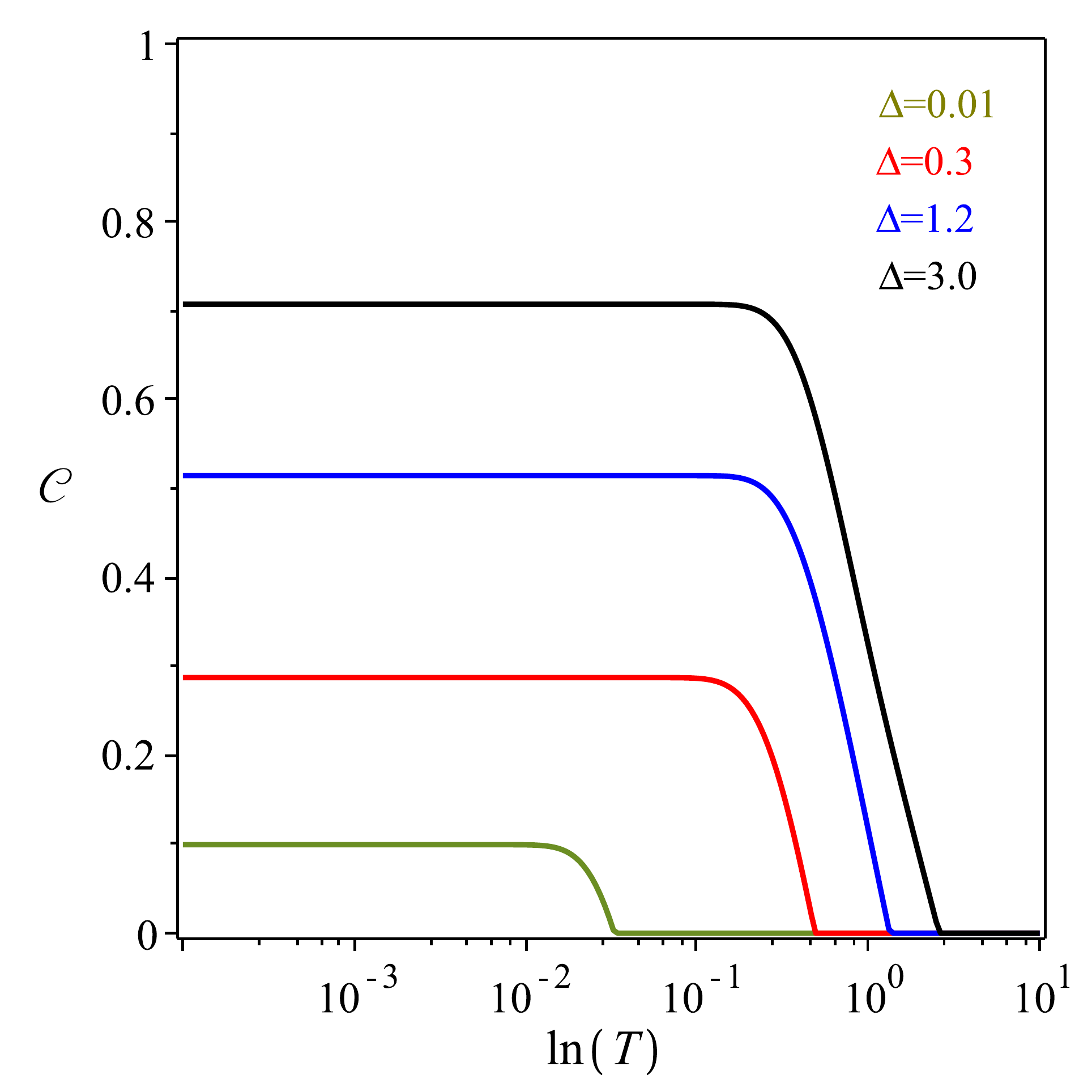}\caption{\label{concurrence:fixed_w:D_eq_w}(Color online) Concurrence $\mathcal{C}$ as a function
of temperature $T$ in the logarithmic scale, for $w=0.1$ (green
curve), $w=1.0$ (red curve), $w=2.0$ (blue curve), $w=3.0$ (black
curve) and different values of $\Delta$. It is assumed that $\Delta < w$ in each case.}
\end{figure}

In Figure \ref{concurrence:fixed_w:D_eq_w}, the concurrence $\mathcal{C}$ is shown as a function of temperature $T$ in the logarithmic scale, at a fixed value of $\Delta=0.01$ and $w=0.1$ (green
curve), $\Delta=0.3$ and $w=1.0$ (red curve), $\Delta=1.2$ and $w=2.0$ (blue curve), $\Delta=3.0$ and $w=3.0$
(black curve).  Initially, one can observe that, when $\Delta<w$, the concurrence is weak $\mathcal{C}\approx 0.1$,  (green curve), $\mathcal{C}\approx 0.28$ (red curve) and  $\mathcal{C}\approx 0.51$ (blue curve). In addition, it can be observed that the concurrence decreases with the increasing of temperature $T$. On the other hand, when we consider $w=\Delta$, the entanglement is more robust at low temperatures $\mathcal{C}=\frac{\sqrt{2}}{2}$ (black curve). By analyzing the matrix elements of the density operator, it is observed that the only single non-zero state vector in this case is given by  $|\varphi_{3}\rangle=  \frac{1}{\sqrt{2}\sqrt{2+\sqrt{2}}}\left(|11\rangle+(1+\sqrt{2})|00\rangle\right)$, at $T=0$. These results are in accordance with expectations, since it is showing that the larger the masses of the state for fixed distances (determined by $\Delta$) and energy scale of the model (determined by $w$), the more intense is the gravity-mediated entanglement, and a similar behavior is found for smaller distances between the particles. However, most interestingly, it shows that the threshold temperature, from which the entanglement vanishes also increases with $\Delta$ and $w$, shall contribute to describe the balance between the needed temperature, mass, distance and energy scales in the type of experiment that aims to measure this effect, which is fundamental for the study of the interplay between gravity and quantum mechanics. However, we further discuss this issue in more realistic scenarios that we address later on in the paper.

\begin{figure}
\centering
\includegraphics[scale=0.4]{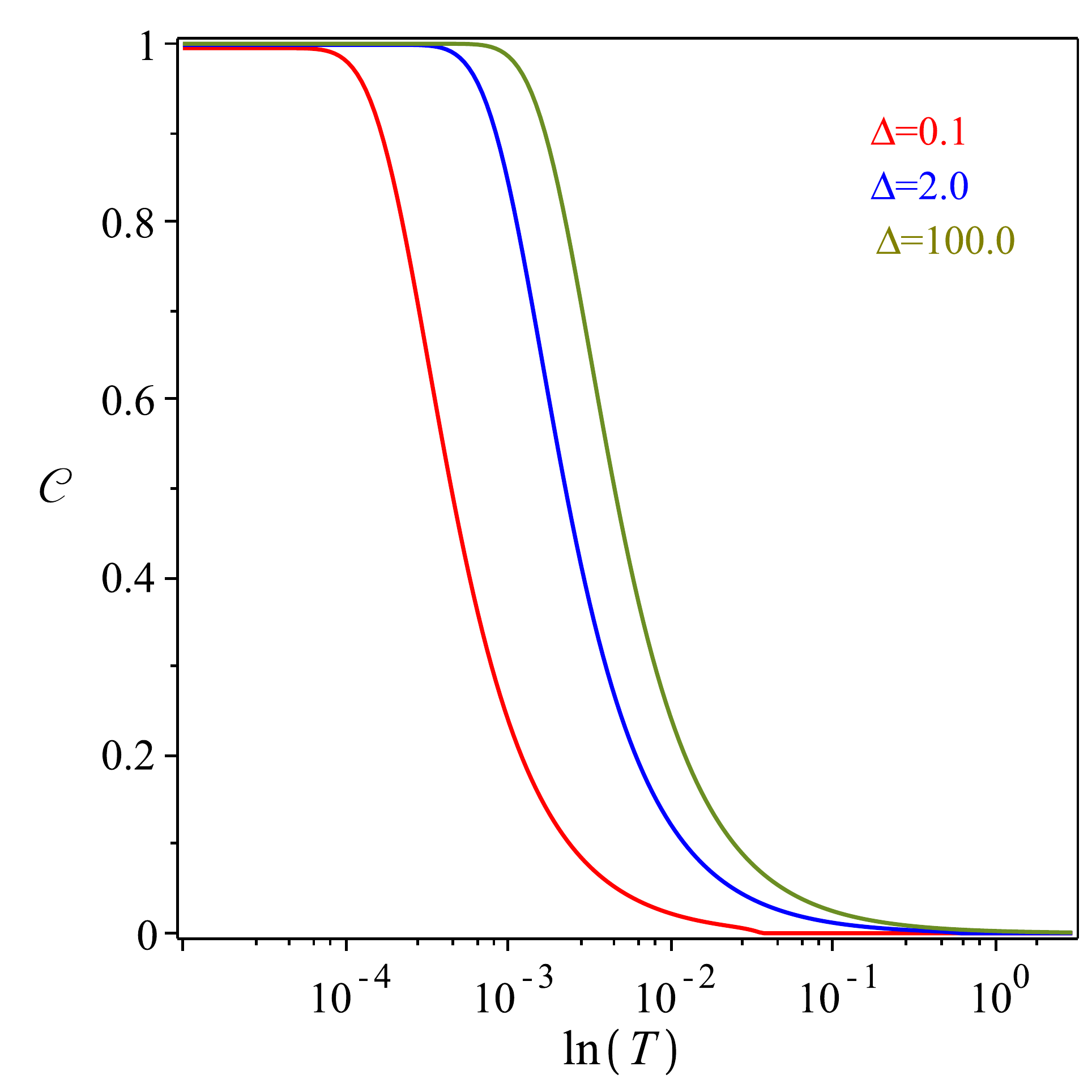}\caption{\label{concurrence:variable_w:D_geq_w}(Color online) Concurrence $\mathcal{C}$ as a function of $T$ in the logarithmic scale, for $w=0.01$ (red curve), $w=0.1$ (blue curve), $w=1.0$ (green curve) and different values of $\Delta$. It is assumed that $\Delta > w$ in each case.}
\end{figure}

In Figure \ref{concurrence:variable_w:D_geq_w}, assuming $\Delta>w$, we observe a significant increase  in the concurrence. In fact, the concurrence reaches its maximum value when the Newton potential parameter $\Delta$ significantly increases in comparison to the energy parameter $w$. However, as the temperature rises, the concurrence decreases until it finally disappears. This condition for maximizing the concurrence can be useful for the planning of experiments, which although not following exactly this kind of model, presents some similarities, from which this analysis may be proved useful.

\begin{figure}
\centering
\includegraphics[scale=0.4]{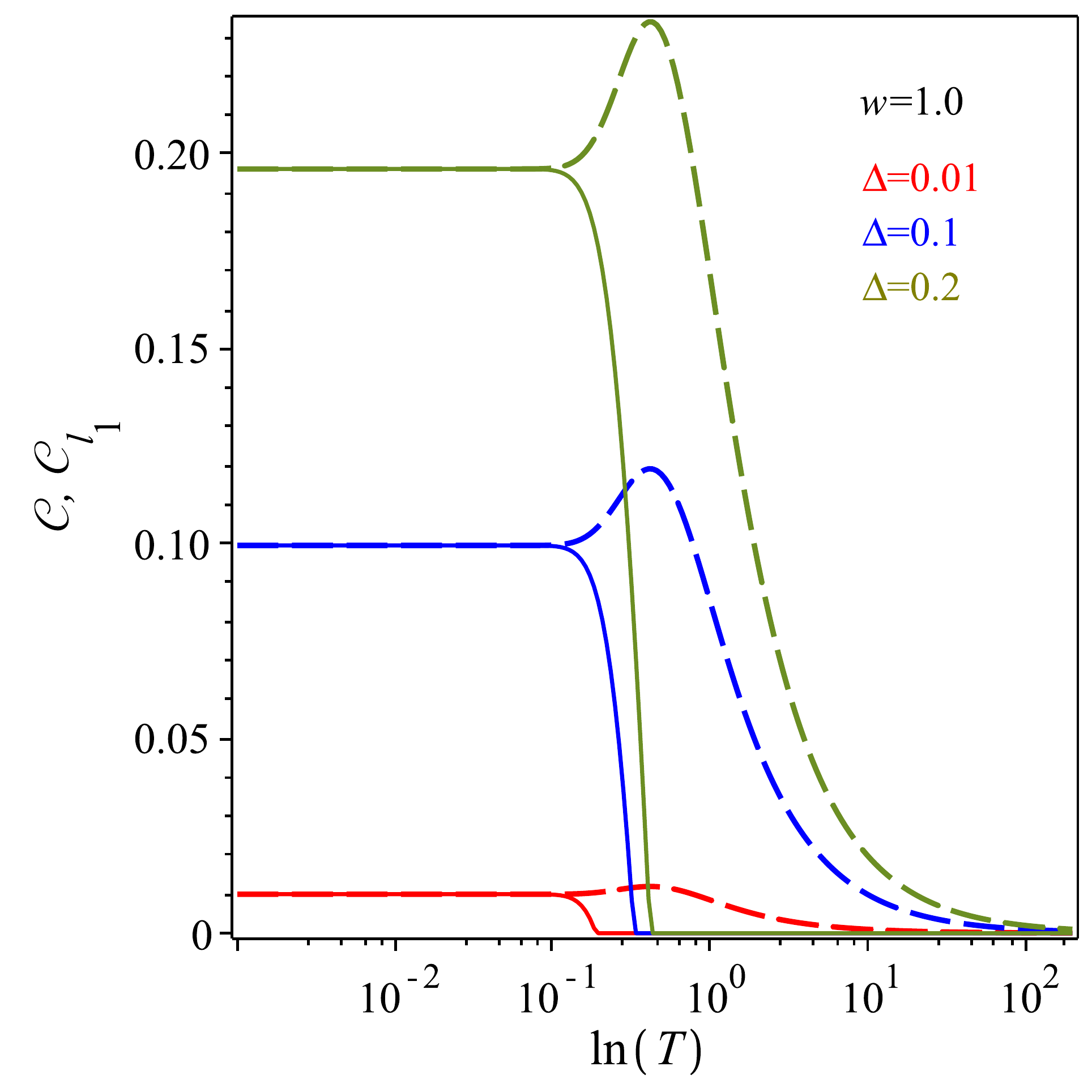}\caption{\label{concurrence_and_coherence:fixed_w:D_leq_w}(Color online) Concurrence $\mathcal{C}$ (solid curve) and the quantum
coherence $\mathcal{C}_{l_{1}}$ (dashed curve) as a function of $T$
in the logarithmic scale for parameter set $\Delta=0.01$ (red curve),
$\Delta=0.1$ (blue curve) and $\Delta=0.2$ (green curve) and fixed
value $w=1.0$. In each case, it is assumed that $\Delta < w$.}
\end{figure}

The $l_{1}$-norm of coherence $\mathcal{C}_{l_{1}}$ is an important
quantity to study quantum coherence. In order to investigate the relation
of entanglement and quantum coherence, here we consider the concurrence
and $l_{1}$-norm. In Figure \ref{concurrence_and_coherence:fixed_w:D_leq_w}, we illustrate the concurrence
$\mathcal{C}$ and $\mathcal{C}_{l_{1}}$ as a function of $T$ in
the logarithmic scale for $\Delta<w$. In particular here, we consider $w=1$ and $\Delta=\{0.01,0.1,0.2\}$. We note that, for low temperatures, there is an overlap between the $l_{1}$-norm of coherence $\mathcal{C}_{l_{1}}$
(dashed curve) and the thermal entanglement (solid curve). This is
due to the fact that the quantum coherence and the thermal entanglement have the same order of magnitude in the low temperature regime, i.e., the quantum coherence is composed
solely of non-local correlations, which is consistent with the previous
result \cite{tan}. As the temperature increases, thermal
fluctuations generate an increasing of the quantum coherence, while the thermal
entanglement decreases until it vanishes at a threshold temperature; these fluctuations raise non-entangled quantum correlations, i.e., quantum discord \cite{tan}. This jump is working as a siren that marks the threshold temperature from which entanglement is no longer observable. On the other hand, it is clear from Figure \ref{concurrence_and_coherence:fixed_w:D_leq_w} that, when we increase the Newton potential $\Delta$, the quantum correlations are more robust.

\begin{figure}
\centering
\includegraphics[scale=0.4]{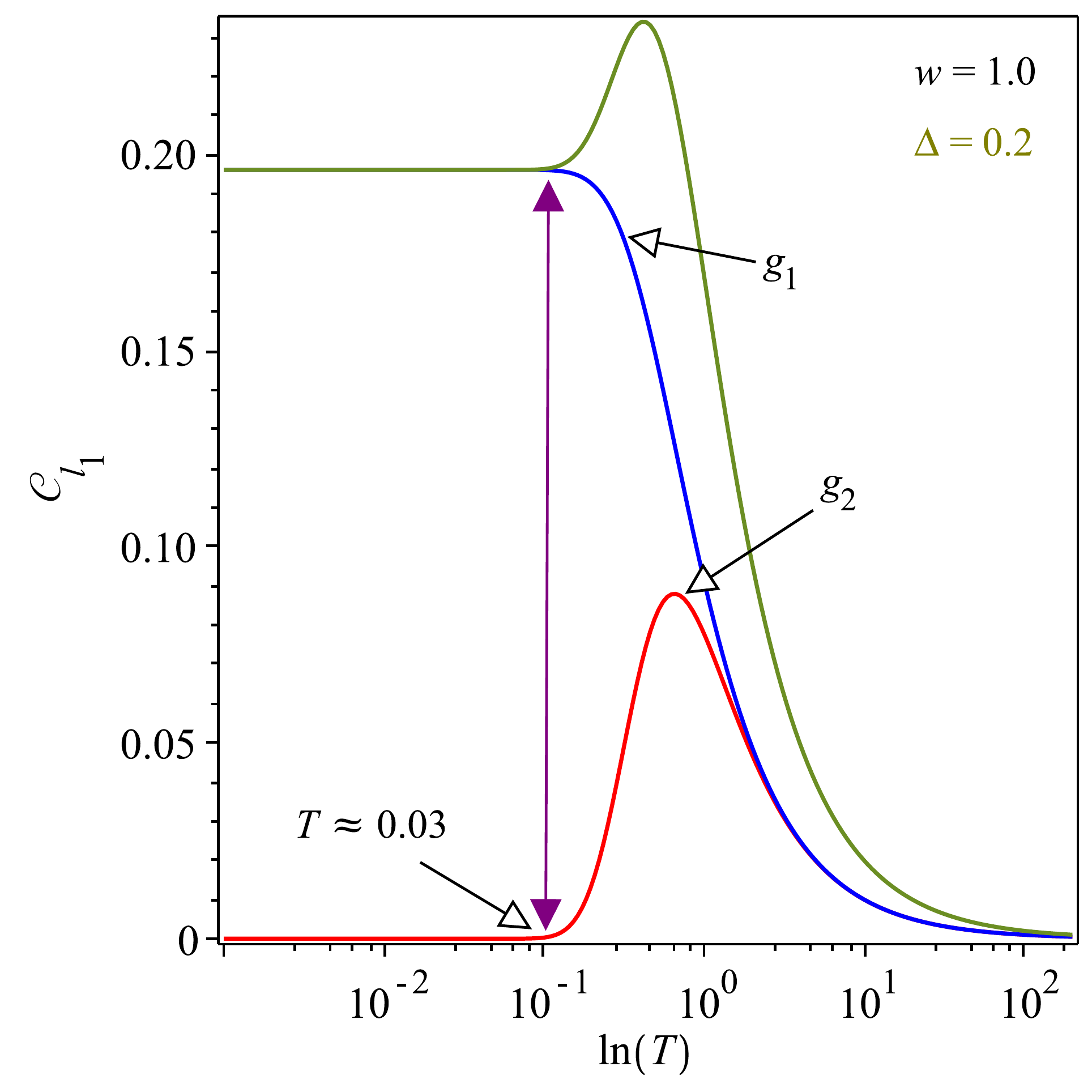}\caption{\label{comparation} (Color online) Quantum coherence $\mathcal{C}_{l_{1}}$ (green curve), $g_{1}$ (blue curve), $g_{2}$ (red curve) as a function of $T$ in the logarithmic scale for parameter set $w=1.0$ and $\Delta=0.2$. The magenta vertical solid line indicates the threshold temperature from which thermal fluctuations occur.}
\end{figure}

In order to understand the behavior of the $l_{1}$-norm, we rewrite Equation (\ref{eq:norm}) as $\mathcal{C}_{l_{1}}=g_{1}+g_{2}$, where $g_{1}=2|\rho_{1,4}|$ and $g_{2}=2|\rho_{2,3}|$. Note that the only term of $l_{1}$-norm that depends on the ground state $\varepsilon_{3}$ is $g_{1}$. In Figure \ref{comparation}, we observed that $\mathcal{C}_{l_{1}}\approx g_{1}$ for very low temperature, at $T \approx 0.03$, the term $g_{1}$ starts to decay, whereas the term $g_{2}$ is initially null from $T=0$ to $T \approx 0.03$, so, by thermal fluctuations, it increases at a rate greater than decay of the term $g_{1}$. This way, the quantum coherence reaches the maximum ($\approx$0.233) at $T\approx 0.433$, and then decreases monotonically. This shows that temperature can generate non-entangled quantum correlations.

\begin{figure}
\centering
\includegraphics[scale=0.4]{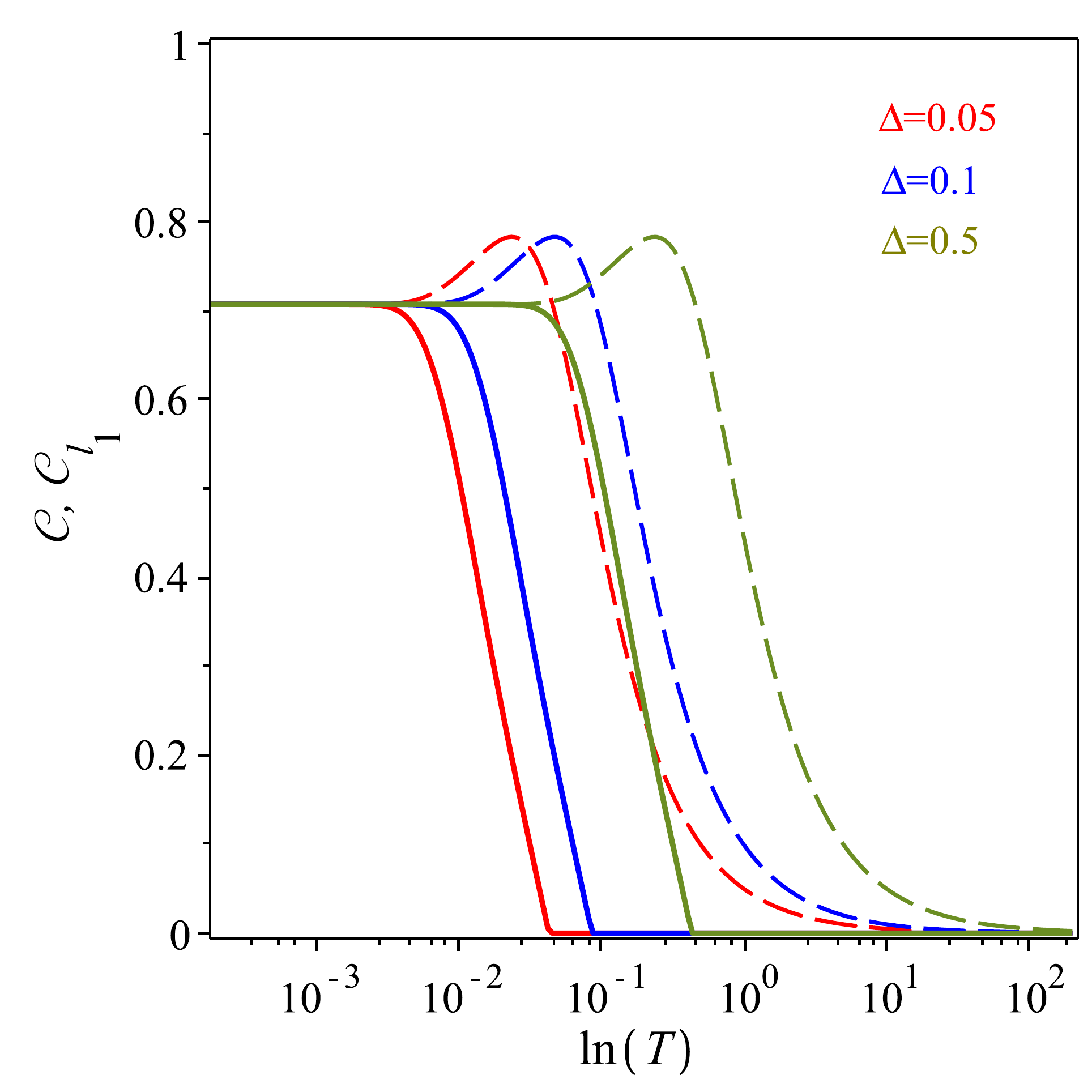}\caption{\label{concurrence_and_coherence:fixed_w:D_eq_w}(Color online) Concurrence $\mathcal{C}$ (solid curve) and the quantum coherence $\mathcal{C}_{l_{1}}$ (dashed curve) as a function of $T$
in the logarithmic scale for parameter set $w=0.05$ and $\Delta=0.05$
(red curve), $w=0.1$ and $\Delta=0.1$ (blue curve), $w=0.5$ and
$\Delta=0.5$ (green curve). Here, it is assumed that $\Delta = w$.}
\end{figure}

In Figure \ref{concurrence_and_coherence:fixed_w:D_eq_w}, another plot of $\mathcal{C}_{l_{1}}$ norm of coherence is shown as a function of the temperature $T$ in the logarithmic scale.  In this figure, we compare the thermal quantum correlations in three cases, namely $\Delta=\{0.05, 0.1, 0.5\}$. For all cases, we consider $\Delta=w$. This condition means that the difference in the energy scale due to the gravitational interaction is equal to the energy difference of each individual two-level system. As can be seen, the thermal quantum coherence for low temperatures is $\mathcal{C}=\mathcal{C}_{l_{1}}=\frac{\sqrt{2}}{2}$. As the temperature increases, thermal fluctuations are introduced into the system; this should generate an increment in non-entangled quantum correlations, signaled by a growth in quantum coherence, which then shortly afterwards decreases monotonically as the temperature increases.

\begin{figure}
\centering
\includegraphics[scale=0.4]{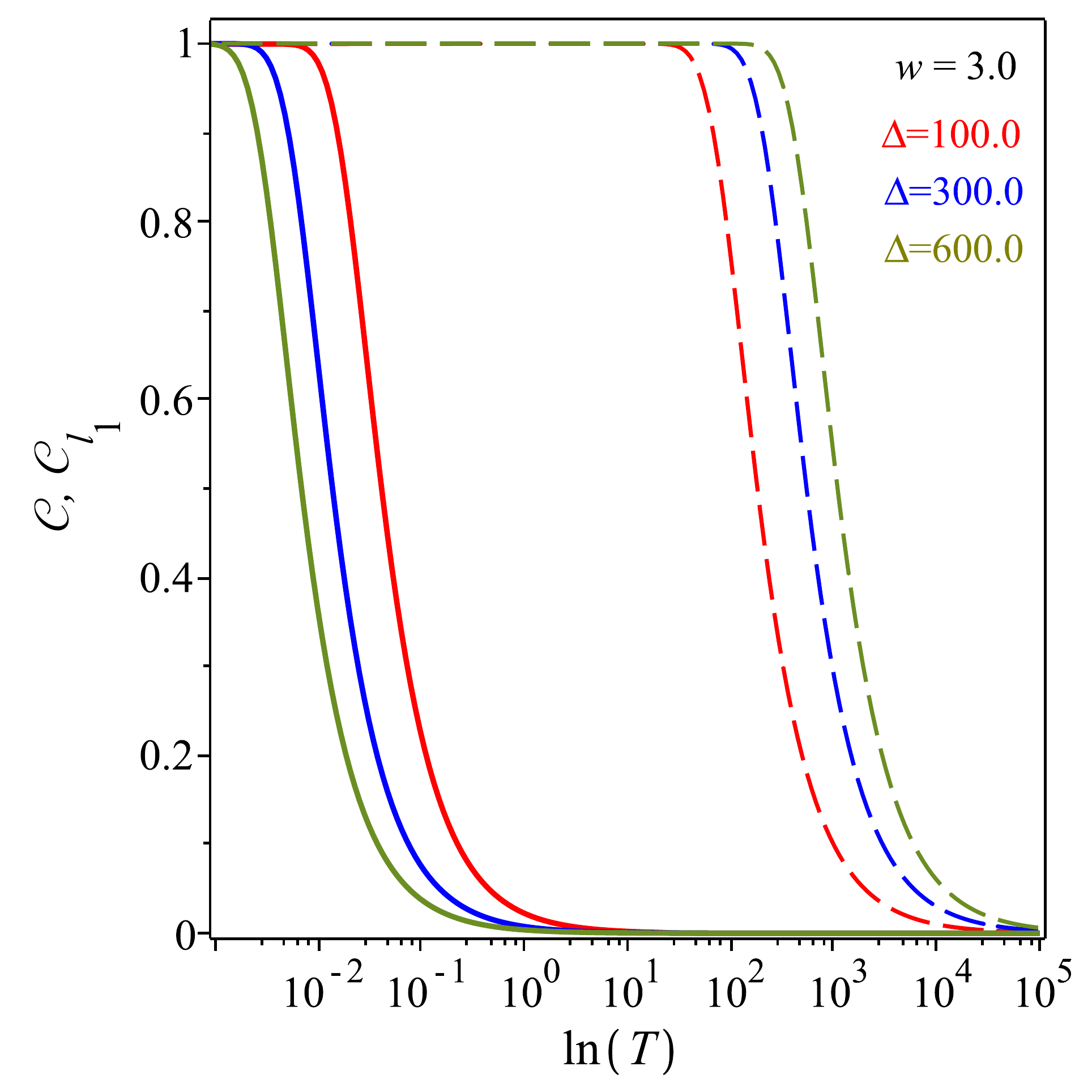}\caption{\label{concurrence_and_coherence:fixed_w:D_geq_w}(Color online) Concurrence $\mathcal{C}$ (solid curve) and the quantum coherence
$\mathcal{C}_{l_{1}}$ (dashed curve) as a function of $T$ in the logarithmic scale with fixed values $w=3.0$, $\Delta=100$, $\Delta=300$ and $\Delta=600$. It is assumed that $\Delta > w$ in each case.}
\end{figure}

In Figure \ref{concurrence_and_coherence:fixed_w:D_geq_w}, we compare the behavior of the concurrence and quantum coherence versus the temperature in the logarithmic scale for different values of the Newton potential $\Delta$ and assuming a fixed value of the energy parameter $w=3.0$; here, we consider $\Delta>w$. It is easy to see a significant increase in the thermal quantum correlations; the entanglement is $\mathcal{C} \approx 1$ for $\Delta>w$ ($\mathcal{C}=1$ when $\Delta \gg w$) for low temperatures. On the other
hand, the $C_{l_1}$ coherence norm shows a significant increase in the non-entangled
quantum correlations regarding low temperatures. However, we notice that, for the conditions assumed on the determination of $w$ and hierarchy between the parameters $(\Delta,w)$, we see that, although the concurrence starts fading at lower temperatures as $\Delta$ increases, the opposite behavior occurs for the coherence. We stress the importance of achieving this kind of condition (energy scale of the gravitational interaction higher than the energy scale of the individual two-level system) in order to enhance the capability of detecting this phenomenon.


\begin{figure}
\centering
\includegraphics[scale=0.4]{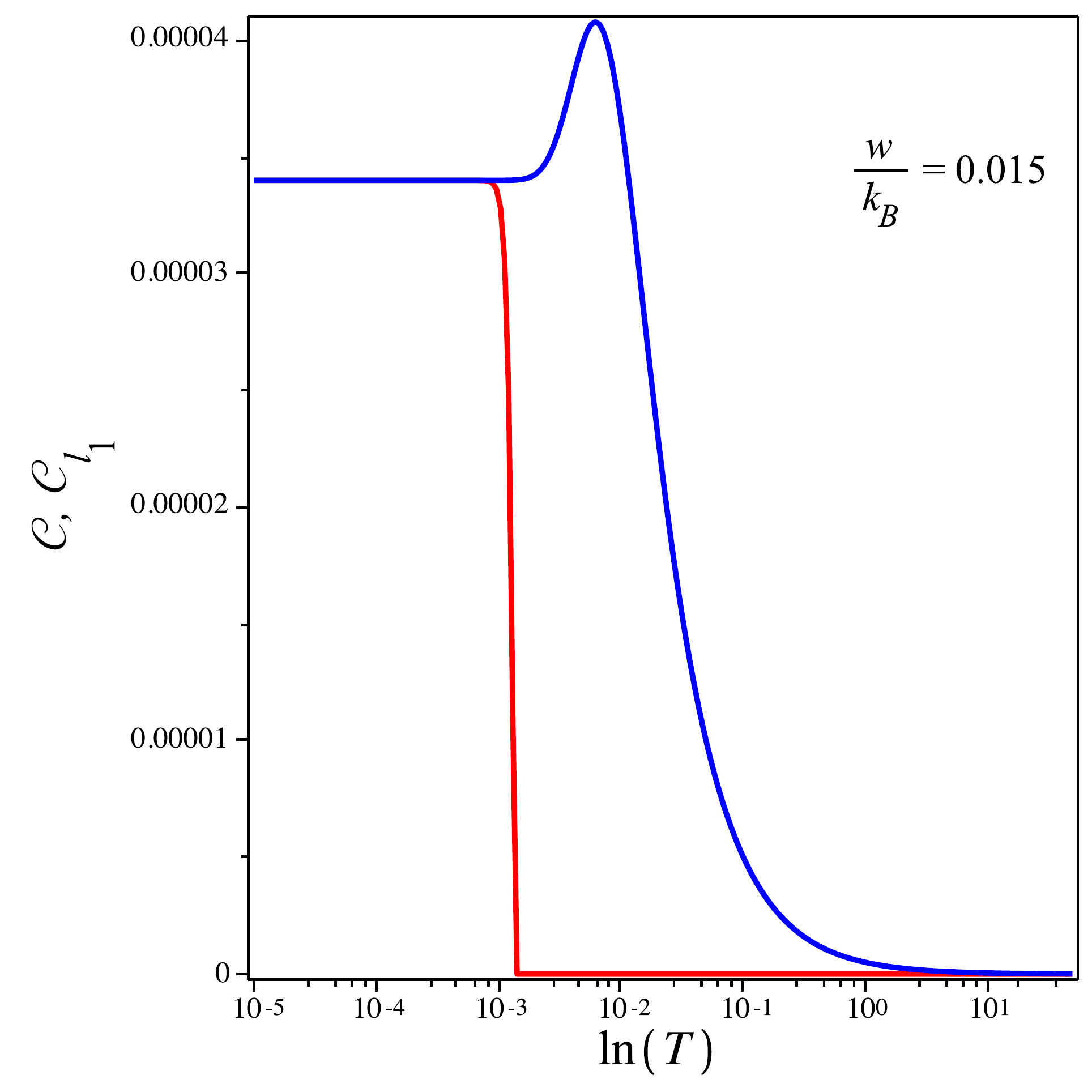}\caption{\label{concurrence_and_coherence:fixed_w:Marletto}(Color online) The plot shows the concurrence $\mathcal{C}$ (red curve) and the quantum
coherence $\mathcal{C}_{l_{1}}$ (blue curve) as a function of $T$ in the logarithmic scale, assuming $\Delta/k_B=0.5101\times 10^{-6}$. This case lies within the $\Delta < w$ class.}
\end{figure}

In the following, we will study the thermal quantum correlations for the parameters suggested in \cite{Marletto:2017kzi,kris}. First, we analyze the thermal entanglement and quantum coherence, namely the $\mathcal{C}_{l_{1}}$ norm, with two mesoscopic test masses $m_{1}\approx m_{2}\approx10^{-12}\, \text{kg}$ separated by $d=1.0\, \mu \text{m}$  (see \cite{Marletto:2017kzi}) and $L=\frac{d}{2}$.  In what follows, we shall recover the Boltzmann constant $k_B\approx 1.4\times 10^{-23}\, \text{J}\cdot \text{K}^{-1}$. In Figure \ref{concurrence_and_coherence:fixed_w:Marletto}, we illustrate the thermal entanglement and quantum coherence as a function of the temperature in the logarithmic scale with $\frac{w}{k_{B}}=0.015$ and $\frac{\Delta}{k_{B}}=0.5101\times10^{-6}$.  From the figure, one can see that the concurrence (red curve) is significantly weak $(\mathcal{C}\approx  3.4\times10^{-5})$ at $T=0\, \text{K}$. Then, the thermal entanglement monotonically decreases with increasing temperature until it rapidly drops to the zero value at the threshold temperature $T_{th}\approx 0.0013658\, \text{K}$.  On the other hand, we can see in the figure that the quantum coherence (blue curve) is equal to the concurrence at the low temperature region, that is, the quantum coherence captures the total thermal entanglement. Soon afterwards, as the temperature increases, the thermal fluctuations generate an increased quantum coherence, while the thermal entanglement decreases until it vanishes at the threshold temperature. Finally, the quantum coherence monotonously fades out. This scenario was also shown in Figure \ref{concurrence_and_coherence:fixed_w:D_leq_w}, and the qualitative discussion that took place above also applies here. However, quantitatively, we present the threshold temperature for which the entanglement survives in this scenario, which might contribute to the design of experiments to test this potential quantum aspect of gravity.

\begin{figure}
\centering
\includegraphics[scale=0.4]{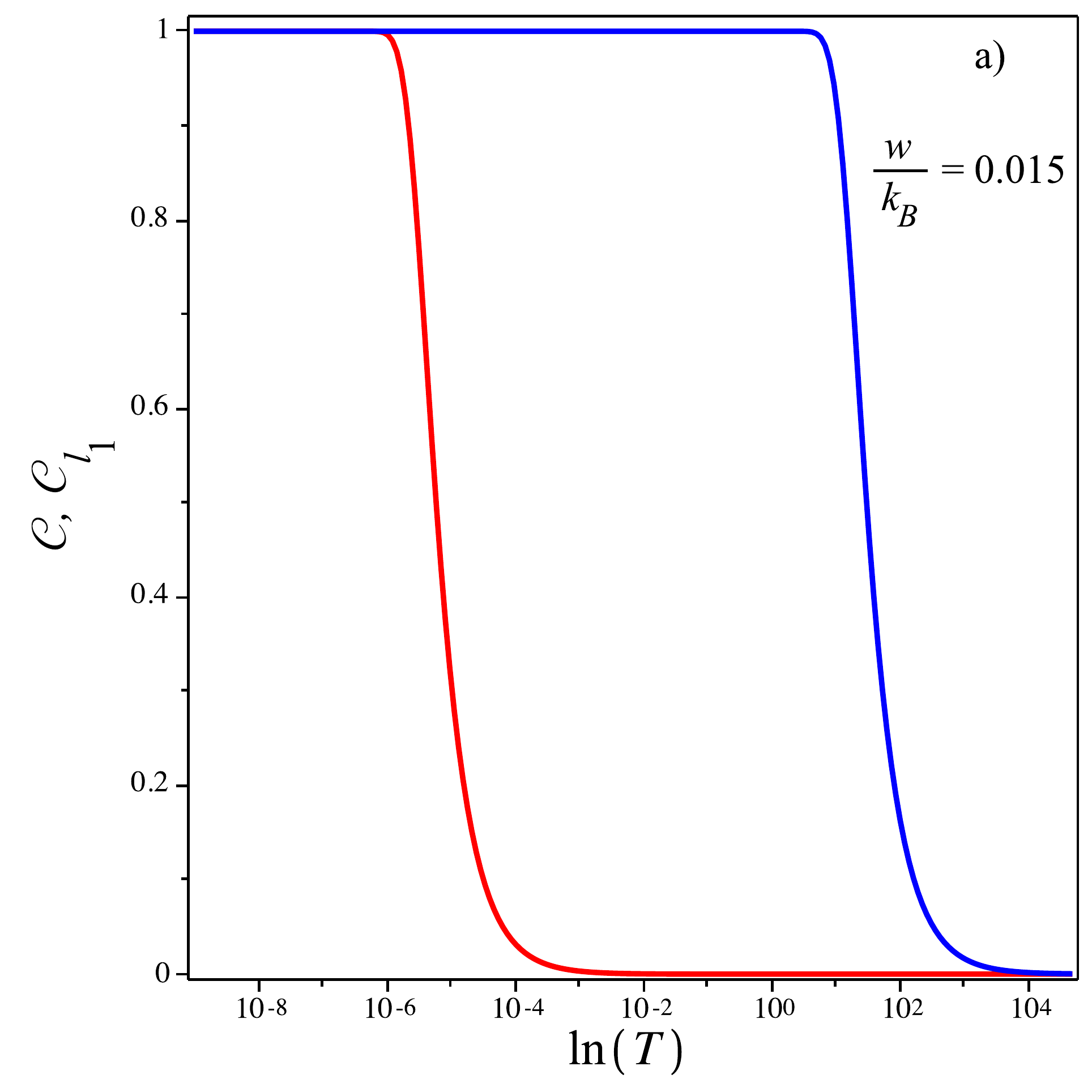}
\includegraphics[scale=0.4]{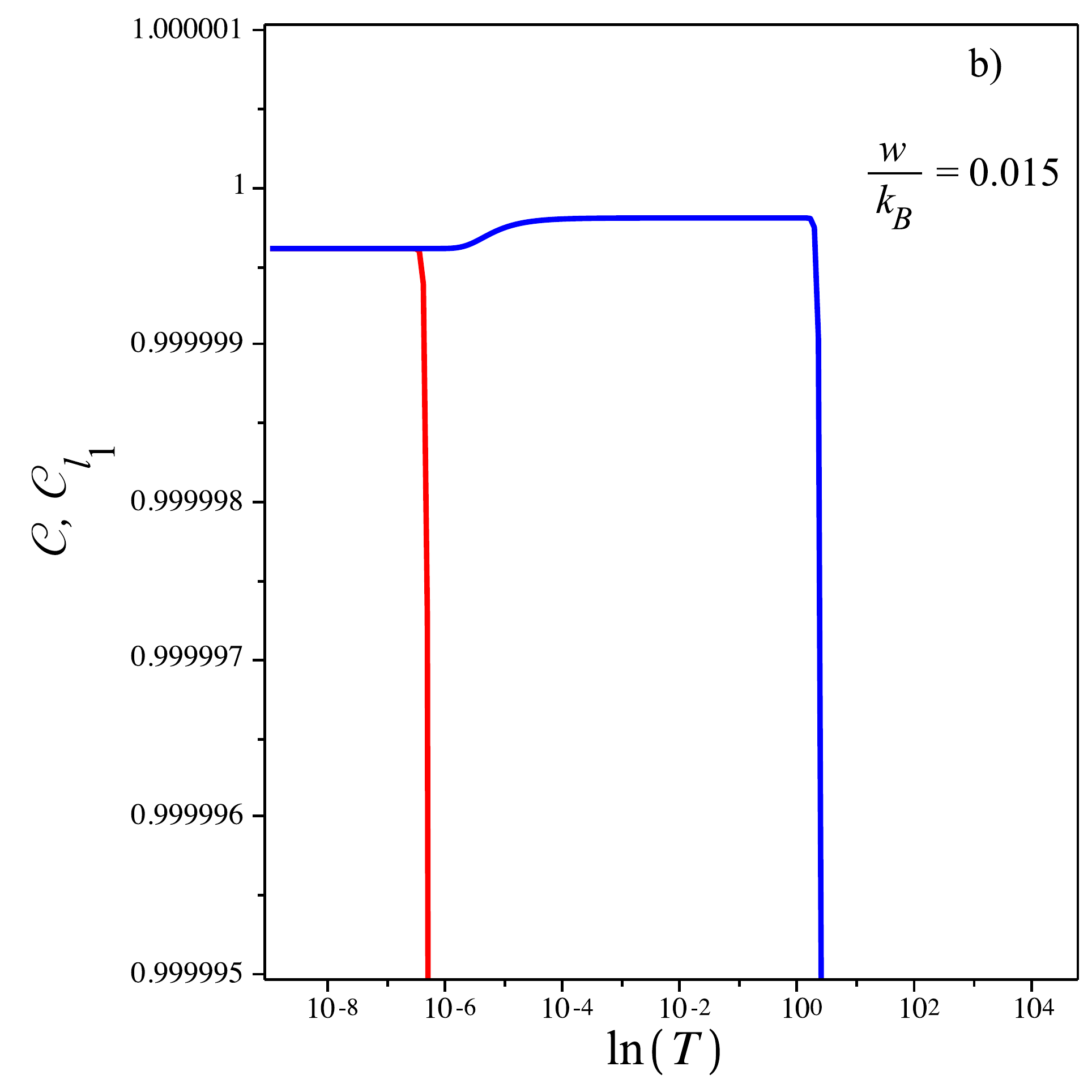}
\caption{\label{concurrence_and_coherence:fixed_w:Kris}(Color online) The plot shows the concurrence $\mathcal{C}$ (red curve) and the quantum
coherence $\mathcal{C}_{l_{1}}$ (blue curve) as a function of $T$, assuming $\Delta/k_B=17.0072$. This case lies within the $\Delta > w$ class.}
\end{figure}

The second setup we consider was recently proposed by Krisnanda et al. \cite{kris}. Here, we consider the masses $m_{1}\approx m_{2}\approx10^{-7}\, \text{kg}$ separated by $d=0.3\, \text{mm}$ and $L=\frac{d}{2}$. The configuration is schematically depicted in Figure \ref{concurrence_and_coherence:fixed_w:Kris} with $\frac{\Delta}{k_{B}}=17.0072$,  $\frac{w}{k_{B}}=0.015$. Our results show that the curves for the concurrence $\mathcal{C}$ and quantum coherence $\mathcal{C}_{l_{1}}$ behave similarly to the Figure \ref{concurrence_and_coherence:fixed_w:Marletto}, but the concurrence in this case is $\mathcal{C}\approx  0.9999996$ at $T=0\, \text{K}$. However, with an increase of the temperature, suddenly the concurrence decreases until it disappears close to the threshold temperature at $T_{th}\approx 2.485053\, \text{K}$ (see Figure \ref{concurrence_and_coherence:fixed_w:Kris}a). 

To better understand the behavior of quantum coherence after the thermal entanglement  begins to decrease, we have plotted the entanglement and quantum coherence vs. temperature in Figure \ref{concurrence_and_coherence:fixed_w:Kris}b. We can see clearly from the figure that the $l_{1}$-norm slightly increases due to the thermal fluctuations and then decreases monotonically as soon as temperature $T$ increases. Here, we verify a strong increment in initial quantum correlations due to the the increase in the intensity of the gravitational interaction in comparison to the previous case. This condition also allows one to have a higher threshold temperature, from which entanglement drops. This also shows the difficulties for experimentally verifying this phenomenon, since the temperature range in which the systems remaining entangled depends significantly on their geometric/gravitational parameters, whose increment constitutes a challenge by themselves.


\section{Conclusions}\label{sec:conc}
In this article, we have studied the effect of temperature produced by a thermal bath in the entanglement and quantum coherence induced by the gravitational interaction of a system of two massive particles confined in two distinct double-well potentials.
\par
We verified the decay of these quantum correlations with the growth of the thermal bath's temperature. For low temperatures, the quantum correlations described by the quantum coherence coincide with the entanglement, i.e., all quantum correlations are entangled in this regime; this is due to the form of the thermal density operator of our model and is consistent with results found in \cite{tan}. Following the decay of the concurrence, the $l_1$-norm slightly grows due to thermal fluctuations, which afterwards monotonically decays with the growth of the temperature.
\par
We have also analyzed the thermal effects on these quantifiers for the parameters suggested in the setups of Marletto et al. \cite{Marletto:2017kzi} and Krisnanda et al. \cite{kris}. In the first setup, our results show that the concurrence and $l_1$-norm of quantum coherence is particularly weak assuming such parameters, but non-null, thus demonstrating the existence of thermal entanglement between massive particles mediated by the Newtonian interaction in a thermal bath model. In the second setup, the thermal concurrence and $l_{1}$-norm of quantum coherence turn out to be robust against decoherence for low temperatures. This is due to fact that the latter case proposes states with masses that are five orders of magnitude larger than the former. It is also interesting to verify that entanglement reaches almost maximal values when $\Delta > w$ at low temperatures, which was also observed in previous cases analyzed in this paper.
\par
This analysis can impact modern searches for traces of the quantum nature of the gravitational field using underground or tabletop experiments, since we showed the presence of a characteristic threshold temperature from which the system loses its entangled nature and could no longer be a laboratory for witnessing the quantum nature of gravitational interaction. Besides that, we verified conditions that enhance quantum correlations' quantifiers: the gravitational energy difference between the two cat states, $\Delta$, being larger than the energy level difference of the particles individually, $w$.
\par 
Future explorations of this work could include further thermodynamic applications of this system and the investigation of whether this behavior persists in more complicated set ups.










\section*{Acknowledgements}This work was partially supported by CNPq, CAPES, and FAPEMIG. M.R. would like to thank the National Council for Scientific and Technological Development-CNPq grant 317324/2021-7. I.P.L. was partially supported by the National Council for Scientific and Technological Development–CNPq grant 306414/2020-1 and by the grant 3197/2021, Para\'iba State Research Foundation (FAPESQ). I.P.L. would like to acknowledge the contribution of the COST Action CA18108.


\end{document}